\documentclass[onecolumn,noshowpacs,nofootinbib,colorlinks,hyperindex,unicode,10.5pt]{revtex4-2}
\usepackage{graphicx}
\usepackage{feynmp-auto}
\usepackage{tikz-feynman}
\usepackage{amssymb}
\usepackage{color,xcolor}

\usepackage{amsfonts,bm,amsmath,amssymb,tikz,accents}
\usepackage[eulergreek]{sansmath}
\usepackage{indentfirst,bigints}
\usepackage{hyperref,upgreek}
\usepackage{bm,graphics,color}
\usepackage{feynmp-auto}
\usepackage{slashed}
\counterwithout{equation}{section}
\usepackage{hyperref}
\usepackage{pgfplots}
\pgfplotsset{compat=1.16}

\hypersetup{hidelinks,colorlinks=true,breaklinks=true,urlcolor= blue}
\hypersetup{%
  colorlinks = true,
  linkcolor  = blue,
  citecolor = cyan,
}
\usepackage{textgreek}
\usepackage{color,xcolor}
\newcommand{\gdualn}[1]{\overset{\:{}^{{}^{\boldsymbol{\neg}}}}{\smash[t]{#1}}} %Elko dual

\def\0{\mbox{\boldmath$\displaystyle\mathbb{O}$}}

\def\s{\mbox{\boldmath$\displaystyle\boldsymbol{\sigma}$}}

\def\p{\mbox{\boldmath$\displaystyle\boldsymbol{p}$}}

\newcommand{\orcidicon}{%
	\begin{tikzpicture}
	\draw[lime, fill=lime] (0,0)
		circle [radius=0.16]
		node[white] {{\fontfamily{qag}\selectfont \tiny ID}};
	\draw[white, fill=white] (-0.0625,0.095)
		circle [radius=0.007];
	\end{tikzpicture}	\hspace{-2mm}
}
\newcommand\orcidg{{\href{https://orcid.org/0000-0002-7942-7941}{\orcidicon}}}
\newcommand\orcidRR{{\href{https://orcid.org/0000-0002-8283-2577}{\orcidicon}}}
\newcommand\orcidRdR{{\href{https://orcid.org/0000-0003-3978-532X}{\orcidicon}}}
\newcommand\orcidCYL{{\href{https://orcid.org/0000-0002-6926-2717}{\orcidicon}}}

\def\nn{\nonumber }
\usepackage{float,xcolor,upgreek}
\usepackage{color} 
\usepackage{tikz-feynman}
\tikzfeynmanset{compat=1.0.0}
\newcommand{\beq}{\begin{eqnarray}}
\newcommand{\eeq}{\end{eqnarray}}
\newcommand{\bea}{\begin{eqnarray}}
\newcommand{\eea}{\end{eqnarray}}

\begin{document}

\title{On Wigner Degeneracy in Elko theory: Hermiticity and Dark Matter}

\author{Gabriel Brand\~ao de Gracia\orcidg{}}
\affiliation{Federal University of ABC, Center of Mathematics, Santo Andr\'e, 09210-580, Brazil.}
\email{g.gracia@ufabc.edu.br}
\email{roldao.rocha@ufabc.edu.br}

\author{Rold\~ao da Rocha\orcidRdR{}}
\affiliation{Federal University of ABC, Center of Mathematics, Santo Andr\'e, 09210-580, Brazil.}

\author{Rodolfo Jos\'e Bueno Rogerio\orcidRR{}}
\affiliation{Centro Universitário UNIFAAT, Atibaia-SP, 12954-070, Brazil.}
\email{rodolforogerio@gmail.com}

\author{Cheng-Yang Lee\orcidCYL{}}
\affiliation{Center for Theoretical Physics, College of Physics, Sichuan University, Chengdu, 610064, China.}
\email{cylee@scu.edu.cn}

%%%%%% Abstract %%%%%%

\begin{abstract}
\indent  In this paper, we provide a set of Hermitian interactions for quantum fields based on Elko, considering the recent achievements concerning the most general form of singular spinors and Wigner degeneracy. We consider Hermiticity and renormalizability a criterion to define the derivative Elko-Higgs interaction as the suitable candidate for a dark coupling. Then, the free parameters of the model are fixed by cosmological constraints on the dark matter abundance and limits on the electron-dark matter scattering mediated by the Higgs.\end{abstract}

\maketitle

%%%%%% Main Text %%%%%%

\section{Introduction}
The analysis of a wide set of galaxy rotation curves and
gravitational lensing effects  associated with their clusters
indicates that there exists a type of matter beyond the observed luminous one \cite{Persic:1995ru,Clowe:2006eq}. The mass profiles of galaxies, computed using mass-to-light ratios in the stellar disks and stellar distributions in spirals, are incompatible with the mass profile obtained from the observation of rotation curves. 
This conundrum can be solved by assuming the existence of dark matter (DM) distributed from the galaxy center out to the galactic halo.  DM comprises 26.8\% of the mass-energy content of the universe, which is dominant compared to the baryonic content which reckons for about 5\% of the entire energy density.
DM forged the development and the settling 
of cosmic structures by way of gravitational instability. When the observed distribution in large
galaxies redshift surveys is compared to computational simulations regarding the large structure formation, one concludes that any fundamental particle 
comprising DM is non-relativistic, 
from the moment it had ceased scattering in the
early universe, motivating the cold DM (CDM) scenario \cite{Planck:2018vyg}. This observational corroboration yielded the $\Lambda$CDM model.  DM possibly undergoes a very weak coupling with the photon and all kinds of ordinary matter. These
features explain the difficulty in measuring DM, whose most relevant phenomenological signatures come from the  gravitational sector. 
%The current cosmological model incorporating DM is the so-called $\uplambda CDM$ concerning a cold DM scenario. 
%It means that the DM particles are assumed to be non-relativistic. The associated abundance is estimated to be of order $\Omega_{CDM} \approx 0.26 $ ?????
% . \\
% \indent 
Beyond cosmological and galactic scales, there is also recent interest in laboratory detection such as in nucleon recoil experiments, such as the XENON-nT \cite{XENON:2022ltv,Sassi:2022njl} and the LUX-ZEPLIN \cite{LZ:2022lsv}. These laboratory searches motivate complementary research on the DM microscopic nature as well as its specific particle representation.  

One of the great mysteries in physics yet to be unraveled is related to the compositions of DM. There are several unsolved questions regarding the DM composition, whether it can described by any fundamental particles, and which interactions are allowed. 
%These questions so far do not have a concrete answer. Until now, it is not possible to assert categorically what constitutes it; we can only infer some of its effects. 
%Although there is an enormous effort within the scientific community to try to unveil the true composition of DM, this is by no means an easy task. 
Among all the presumed candidates to encompass DM, the theory of mass dimension one fermions based on Elko spinors has gained considerable attention~\cite{jcap, mdobook}. From a formal and  theoretical point of view, based on the seminal works of Wigner~\cite{wigner1, wigner2}, the current formalism provides a mathematically and physically well-defined attempt to fully understand the particle content as implied
by Poincar\'{e} spacetime symmetries.\footnote{Elko is the German acronym for Eigenspinoren des Ladunskonjugationoperators. In English, it is eigenspinor of the charge conjugation operator.} They are prime candidates to describe DM, as the dark properties of these fermions are associated with the fact that their Elko spinor expansion coefficients are eigenspinors of the charge conjugation operator. We will provide a sharper definition of this last claim at suitable points throughout the paper. Namely, the ones concerning the action symmetries and Feynman rules.

Indeed, Elko effective couplings with the Standard Model are suppressed, with few exceptions, such as the Elko-photon neutral tree-level interaction, implemented by a dimensionless coupling with a Pauli-like structure. This kind of interaction produces a typical signature that may be constrained by experiments involving the DM search \cite{NPB2023,Lee:2015sqj,Alves:2014kta,Dias:2010aa,Agarwal:2014oaa,Duarte:2017svd}. This property is currently associated with experimental bounds on interactions between the photon and DM  \cite{Ahluwalia:2022ttu}. The Elko-photon coupling can be experimentally probed in monophoton events at the LHC \cite{Alves:2017joy,Alves:2014qua}. Interestingly, this kind of coupling naturally suppresses Elko-photon Compton-like scattering and pair annihilation in a type of darkness mechanism characterizing this DM candidate~\cite{NPB2023}.

The discovery of Elko has attracted the attention of many physicists toward understanding the formal aspects underlying Elko. It has motivated a wide range of investigations such as their mathematical properties~\cite{rrjhep, jrspin, polarform2020,Elkopolar,chenggeneral,chenglagrangian,Cavalcanti:2014uta}, applications in cosmology~\cite{saulo1,saulo2,saulo3,daRocha:2011yr},  phenomenology in high-energy physics~\cite{NPB2023,Lee:2015sqj,Alves:2014kta,Dias:2010aa,Agarwal:2014oaa,Duarte:2017svd,Ahluwalia:2022ttu,Alves:2017joy,Alves:2014qua,roldaofermionic,julioperturbative,daRocha:2014dla}, and also the implementation of new fermionic fields in several aspects of gauge/gravity duality and quantum approaches to gravity \cite{deBrito:2016qzl,daRocha:2007sd,daRocha:2009gb,Bonora:2015ppa,Bonora:2014dfa,Meert:2018qzk,Meert:2018qzk}.

Among numerous reasons for establishing fermionic fields constructed upon Elko spinors as natural prime candidates for solving the DM problem, two of them must be highlighted. First, the fact that charge conjugation invariance avoids minimal coupling between this spinor field and electromagnetism, as suitably addressed in key discussions along the article.  Another important characteristic of Elko is the inherent mass dimension one nature of the associated quantum field, instead of mass dimension three-halves as usual for the Dirac fields.
%THE TEXT BELOW IS IPSIS LITERIS REPEATED ON LINE 167! The theoretical discovery of these spinors focused the attention of many physicists on better understanding the formal aspects underlying not only Elko spinors but also a plethora of new types of correlated fermionic fields. It motivated a wide set of investigations regarding several aspects such as their mathematical properties \cite{rrjhep, jrspin, polarform2020,Elkopolar,chenggeneral,chenglagrangian,Cavalcanti:2014uta}, applications in cosmology \cite{saulo1,saulo2,saulo3,daRocha:2011yr},  phenomenology in high-energy physics  \cite{roldaofermionic,julioperturbative,Lee:2015sqj}, and also the implementation of new fermionic fields in several aspects of gauge/gravity duality and quantum approaches to gravity. 
This fact is intrinsically related to the equations of motion governing these spinors, as these objects, although satisfying the Klein-Gordon equation, are not governed by the Dirac equation. Moreover, this specific mass dimensionality avoids the participation of this DM  candidate in standard matter doublets, restricting its possible set of interactions. 
% \textcolor{blue}{In this case, the standard local gauge interactions are not allowed.} 
Thus, Elko and the Dirac fields carry intrinsically different possibilities for their interactions through local gauge fields \cite{aaca}, reinforcing the inherent ``darkness'' of the first. Accordingly, it is worth mentioning that fields based on the Elko as expansion coefficients successfully describe the DM halo near the galactic nuclei. Elko can also support the mass-to-light ratios for ultracompact dwarf galaxies and the precise galaxy rotation curves as well \cite{Ahluwalia:2022ttu,Pereira:2017efk,saulo1,saulo2,Pereira:2017efk}. As we will see, the Hermitian and renormalizable interactions concerning these fermions are restricted to a suitable coupling with the Higgs particle \footnote{More specifically, the neutral scalar non-pure gauge perturbation contained in the Higgs doublet around the quartic potential minimum occurring after spontaneous symmetry breaking.}. However, this ensures a rich phenomenology due to the wide variety of interactions involving the Higgs. It is possible to show that it furnishes a natural explanation for the cosmological observations and also the bounds obtained in laboratory searches regarding electron-DM scattering.

In this work, we investigate a Hermitian derivative coupling of the quantum fields based on Elko spinor coefficients with the Higgs field. Since the latter has also a variety of couplings in the electroweak sector, we will prove that it is sufficient to describe a relevant set of results in both cosmological and laboratory scales.  Sec. \ref{2} is devoted to 
 reviewing fundamental  properties of singular spinors, whose most important representative are the Elko, which are eigenstates of the 
 charge conjugation operator. We highlight the recent theoretical improvements based on the full Lorentz covariant structure ensured by considering the double Wigner degeneracy. The introduction of the new dual compatible with this specific set of spinors is also scrutinized. Later, we provide the structure of the mass dimension one fermion based on Elko spinor coefficients. Sec. \ref{3} is dedicated to explicitly demonstrating how this dual structure ensures a Hermitian action for mass dimension one fermions. We also derive a Hermitian interaction with the Higgs boson. Later, in Sec. \ref{5}, a wider variety of Hermitian couplings based on the previous considerations is presented. However, some of them are non-renormalizable, a feature that can be considered as a criterion to disregard them as a fundamental interaction of nature. However, they can still define interesting effective models paving a robust setup for upcoming research. Therefore, in Sec. \ref{6}  the Feynman rules are derived for our target interaction concerning the Higgs field. In Sec. \ref{7}, the 1-loop radiative corrections are explicitly obtained for this new model to confirm the claims on renormalizability. All the divergent contributions are verified to have such a form that can be absorbed in the renormalization of the bare Lagrangians, as it should be for this kind of theory. The necessity of this explicit verification is highlighted at the beginning of this mentioned section, defining a complementary formal background. To address the central issue of the paper, we first study the M\o ller-like scattering at tree-level in Sec. \ref{8}. We calculate polarized and non-polarized squared amplitudes, revealing positive-definite results, complying with the probabilistic interpretation ensured by the Hermitian structure. Sec. \ref{9} presents a set of processes involving mass dimension one fermions annihilating into Higgs particles, leptons, and vector bosons. These reactions are fundamental for defining DM abundance.  Sec. \ref{10} is associated with the study of the freeze-out process and determining the DM relic. The definition of a CDM scenario, its correlated freeze-out temperature, and cosmological constraints on the free parameters are imposed. In Sec. \ref{12}, we show that the current experimental bounds on DM-electron scattering agree with the previous setting on the parameters of the model. Finally, we conclude in Sec. \ref{13}.

\section{A brief review on Elko and Wigner degeneracy} \label{2}

\indent For the sake of completeness, in this section we review the main aspects of Elko and, more generally,  singular spinors, bearing in mind their relevant properties that will be necessary along the calculations to be performed. These spinors are defined as the ones whose norm under the standard Dirac dual vanishes.  In the general approach to QFT, one considers an arbitrary spinor $\uppsi(p^\mu)$ 
as an object in the $\left(\frac12,0\right)\oplus\left(0,\frac12\right)$ Lorentz group irreducible representation. 
In the Weyl chiral representation, the arbitrary spinor can be represented by \cite{Ahluwalia:2022ttu}
	\begin{equation}	\uppsi(p^\mu) = \binom{		\upphi_{\mathsf{R}}(p^\mu)}{		\upphi_{\mathsf{L}}(p^\mu)}.
\end{equation}
Weyl spinors of right and left chirality transform under the $\left(\frac12,0\right)$ and  the $\left(0,\frac12\right)$ representations, respectively. The distinct  chiral components ${\mathsf{R}}$ and ${\mathsf{L}}$ can be split up from the original spinor considering the chiral projectors, 
\begin{equation} P_{{\mathsf{R}}/{\mathsf{L}}}\equiv \frac{1}{2}\big( \boldsymbol{1}\pm \gamma_5\big).    \end{equation}
 It is worth emphasizing that when one denotes the helicity eigenstates by
\begin{equation}
(\s\cdot\widehat \p) \upphi_\pm(p^\mu) = \pm \upphi_\pm(p^\mu),
\end{equation} where $\boldsymbol{\sigma}$ stands for the Pauli matrices, the 
massless limit of the Dirac equation implies that $\upphi_{{\mathsf{R}}/{\mathsf{L}}}(p^\mu)=\upphi_\pm(p^\mu)$. 
In spherical coordinates, 
\begin{align}
\upphi_+(\boldsymbol{0}) & = \sqrt{\frac{m}{2}}  \left(
									\begin{array}{c}
									\cos\left(\frac{\theta}{2}\right)e^{- i \phi/2}\\
									\sin\left(\frac{\theta}{2}\right)e^{i \phi/2}
								\end{array}
	\right),\qquad\qquad
\upphi_-(\boldsymbol{0}) = \sqrt{\frac{m}{2}}   \left(		\begin{array}{c}
									- \sin\left(\frac{\theta}{2}\right)e^{- i \phi/2}\\
									 \cos\left(\frac{\theta}{2}\right)e^{i \phi/2}
											\end{array}
									\right),
\end{align}
which along the $z$-axis turns into 
\begin{eqnarray}\label{components}
\upphi^{+}(\boldsymbol{0}) = \sqrt{\frac{m}{2}}\left(\begin{array}{c}
1 \\ 
0
\end{array}\right), \qquad\qquad\qquad  \upphi^{-}(\boldsymbol{0}) = \sqrt{\frac{m}{2}}\left(\begin{array}{c}
0 \\ 
1
\end{array}\right),
\end{eqnarray}
With the Wigner time-reversal operator \cite{jcap}
\begin{eqnarray}
\Theta = \left(\begin{array}{cc}
0 & -1 \\ 
1 & 0
\end{array}\right),
\end{eqnarray} 
it paves the possibility to define the following set of singular spinors \footnote{The symbols $\alpha$ and $\beta$ represent trial parameter factors, to be determined under the demand of certain physical requirements to be further explored. The whole domain of the singular spinors is defined by all the possible complex configurations of these parameters.  }
\begin{subequations}
\begin{eqnarray}\label{espi}
\uplambda^S_{\{+,-\}}(\boldsymbol{0})=\left(\begin{array}{c}
\alpha\Theta\upphi^{-*}(\boldsymbol{0}) \\ 
\beta\upphi^{-}(\boldsymbol{0})
\end{array} \right),
\qquad\quad \uplambda^S_{\{-,+\}}(\boldsymbol{0})=\left(\begin{array}{c}
\alpha\Theta\upphi^{+*}(\boldsymbol{0}) \\ 
\beta\upphi^{+}(\boldsymbol{0})
\end{array} \right),
\\\label{nor} 
\uplambda^A_{\{+,-\}}(\boldsymbol{0})=\left(\begin{array}{c}
-\alpha\Theta\upphi^{-*}(\boldsymbol{0}) \\ 
\beta\upphi^{-}(\boldsymbol{0})
\end{array} \right),
\qquad\quad \uplambda^A_{\{-,+\}}(\boldsymbol{0})=\left(\begin{array}{c}
-\alpha\Theta\upphi^{+*}(\boldsymbol{0}) \\ 
\beta\upphi^{+}(\boldsymbol{0})
\end{array} \right),
\end{eqnarray}
\end{subequations}
whereas the following set of degenerated spinors can be proposed\footnote{The original and pivotal aspects regarding degenerated spinors were firstly reported in Appendix B of Ref.  \cite{jcap}.}  \cite{dharamnpb}:
\begin{subequations}
\begin{eqnarray}\label{espi1}
\rho^S_{\{+,-\}}(\textbf{0})&=&\left(\begin{array}{c}
\beta^*\upphi^{+}(\boldsymbol{0}) \\ 
\alpha^*\Theta\upphi^{+*}(\boldsymbol{0})
\end{array} \right),\qquad\quad \rho^S_{\{-,+\}}(\textbf{0})=\left(\begin{array}{c}
\beta^*\upphi^{-}(\boldsymbol{0}) \\ 
\alpha^*\Theta\upphi^{-*}(\boldsymbol{0})
\end{array} \right),\\
\rho^A_{\{+,-\}}(\textbf{0})&=&\left(\begin{array}{c}
-\beta^*\upphi^{+}(\boldsymbol{0}) \\ 
\alpha^*\Theta\upphi^{+*}(\boldsymbol{0})
\end{array} \right),
\qquad\quad \rho^A_{\{-,+\}}(\textbf{0})=\left(\begin{array}{c}
-\beta^*\upphi^{-}(\boldsymbol{0}) \\ 
\alpha^*\Theta\upphi^{-*}(\boldsymbol{0})
\end{array} \right).\label{nor1}
\end{eqnarray}
\end{subequations}
Having defined the rest frame spinors, for an arbitrary momentum they can be written as $\uplambda_h(\textbf{p})= \kappa\uplambda_h(\textbf{0})$ and $\rho_h(\textbf{p})= \kappa\rho_h(\textbf{0})$, acting  the $\left(\frac12,0\right)\oplus\left(0,\frac12\right)$ boost operator
\begin{eqnarray}
\kappa = \sqrt{\frac{E+m}{2m}}\left(\begin{array}{cc}
\boldsymbol{1}+ \frac{\boldsymbol{\sigma\cdot p} }{E+m} & 0 \\ 
0 & \boldsymbol{1}- \frac{\boldsymbol{\sigma\cdot p}}{E+m}
\end{array} \right).
\end{eqnarray} 

Differently from Majorana and Dirac fermions, the kinematical aspects of the singular spinors are regulated by a coupled system of first-order PDEs, mixing the four spinor types under the application of the Dirac kinetic operator. The Elko spinors, defined by $\alpha=i$ and $\beta=1$, are special representatives of this singular class being eigenstates of the charge conjugation operator and, therefore, DM candidates due to their intrinsic neutral nature. Each one of the spinors in Eqs. (\ref{espi}, \ref{nor}) and (\ref{espi1}, \ref{nor1})  satisfy the Klein--Gordon equation. Regarding Elko spinors, they can be divided into two sets of self-conjugate (Eqs. (\ref{espi}, \ref{espi1})) and anti-self-conjugate (Eqs. (\ref{nor}, \ref{nor1})) ones, undergoing the action of the charge conjugation operator. Both eigensets can be later split into two subsets, regarding the sign of the Elko spinor components under the helicity operator \cite{jcap}. The labels $S$ and $A$ are employed to highlight the (anti) self-conjugate  nature of the spinors when the Elko phase is reached by the correct fixation of parameters.

Additionally, one can realize that eight spinors are defined, rather than the four usually defined in the case of Dirac spinors. The degenerated spinors, namely $\rho$, are fundamental ingredients for the theory of singular spinors to  experience locality, Lorentz-invariant spin sums, and Hermitian amplitudes and scatterings. More details can be found in Refs. \cite{dharamnpb,rodolfosingular}.  

To establish a convenient notation, instead of working with eight spinors, one defines the $\xi_h(\boldsymbol{\p})$ spinors as
\begin{subequations}
\begin{eqnarray}
&&\xi_1(\boldsymbol{\p}) = \uplambda^S_{\{+,-\}}(\boldsymbol{\p}), \qquad\qquad\quad \xi_2(\boldsymbol{\p}) = \uplambda^S_{\{-,+\}}(\boldsymbol{\p}),\\
&&\xi_3(\boldsymbol{\p}) = \rho^S_{\{+,-\}}(\boldsymbol{\p}), \qquad\quad\qquad \xi_4(\boldsymbol{\p}) = \rho^S_{\{-,+\}}(\boldsymbol{\p}),
\end{eqnarray}
\end{subequations}
comprising a self-conjugate basis when the system is restricted to the physical Elko phase. It is a manifestation of the Wigner degeneracy associated with the fact that a given eigenspinor of the charge conjugation operator (Elko) flips its eigenvalue if multiplied by the imaginary unit, due to the anti-linear nature of such operator \cite{Ahluwalia:2022ttu,dharamnpb}. 
Similarly, one can define also an anti-self-conjugate basis $\upchi_h(\boldsymbol{\p})$ spinors
\begin{subequations}
\begin{eqnarray}
&&\upchi_1(\boldsymbol{\p}) = \uplambda^A_{\{+,-\}}(\boldsymbol{\p}), \qquad\quad\qquad \upchi_2(\boldsymbol{\p}) = \uplambda^A_{\{-,+\}}(\boldsymbol{\p}),\\
&&\upchi_3(\boldsymbol{\p}) = \rho^A_{\{+,-\}}(\boldsymbol{\p}), \qquad\qquad\quad \upchi_4(\boldsymbol{\p}) = \rho^A_{\{-,+\}}(\boldsymbol{\p}).
\end{eqnarray}
\end{subequations}

We will take a brief pause here to discuss in more detail the dual structure for singular spinors. With the developments following the discovery of Elko spinors and the need for a redefinition of the dual structure, new spinor duals have been playing important roles in the construction of fermionic fields beyond the Standard Model. The obtainment of physical observables from a spinorial theory essentially depends on two features: the trial parameters and the dual structure. As mentioned, considering the Dirac dual structure $\bar\psi = \psi^{\dag}\gamma_0$ for singular spinors, Ref. \cite{daRocha:2005ti}
demonstrated that it leads to vanishing norms for all singular spinor sets, in particular the ones entering the construction of Elko. Therefore, it would prevent the definition of gravitational interactions as well as their detection, unlike what is expected for DM. Moreover, to quantize the associated fields, this dual structure would imply severe issues in the particle interpretation, besides predicting states with negative energy.\\
\indent A manner to circumvent this situation is to search for an alternative dual definition, as initially proposed in \cite{aaca, rjdual}. Thus, one can envisage a new physical scenario with relevant physical information that was previously overlooked. It is possible to state that the most general structure to define the dual is given by $\gdualn{\psi}_{h}(\textbf{p}) = [\mathcal{P}\psi_h(\textbf{p})]^{\dag}\gamma_0$, in which $\mathcal{P} = m^{-1}\gamma^{\mu}p_{\mu}$  stands for the parity operator \cite{speranca}, and $\psi_h(\textbf{p})$ is any arbitrary spinor. The formulation of this spinorial structure can ensure relevant physical observables \cite{beyondlounesto}, invariance/covariance of the physical objects involved \cite{rjdual}, the Takahashi's inversion theorem \cite{rrtaka}, quantum field locality \cite{rodolfonogo}, among other relevant aspects within QFT \cite{rog1,rog2}. 

Thus, based on the recent discussions on spinors and dual structures, its appropriate definition reads 
$\gdualn{\xi}_{h}(\textbf{p}) = [\mathcal{P}\xi_h(\textbf{p})]^{\dag}\gamma_0$ and $\gdualn{\upchi}_{h}(\textbf{p}) = [-\mathcal{P}\upchi_h(\textbf{p})]^{\dag}\gamma_0$. Such a new dual definition sets singular spinors in a well-posed theoretical framework, from both the mathematical and physical points of view. The features mentioned above evince that singular spinors carry irreducible representations of the extended inhomogeneous Lorentz group, as the Dirac and Majorana spinors fields.     

The spinorial structures and the introduced duals imply important results, leading to the following orthonormal relations:
\begin{eqnarray}
&&\gdualn{\xi}_{h}(\p)\xi_{h^{\prime}}(\p) = \frac{m}{2}\left(|\alpha|^2+|\beta|^2\right)\delta_{hh^{\prime}}, \label{ortoF1}
\\
&&\gdualn{\upchi}_{h}(\p)\upchi_{h^{\prime}}(\p) = -\frac{m}{2}\left(|\alpha|^2+|\beta|^2\right)\delta_{hh^{\prime}}. \label{ortoF2}
\end{eqnarray}
Hence, the spin sums read
\begin{eqnarray}
&& \sum_{h} \xi_h(\p)\gdualn{\xi}_{h}(\p)  = \frac{m}{2}(|\alpha|^2+|\beta|^2) \boldsymbol{1},\label{spinsumS}
 \\
&& \sum_{h} \upchi_h(\p)\gdualn{\upchi}_{h}(\p)  = -\frac{m}{2}(|\alpha|^2+|\beta|^2) \boldsymbol{1},\label{spinsumFA}
\end{eqnarray}
which are real and Lorentz covariant. Again we emphasize the importance of the well-defined dual structure and the basis in compliance with Wigner degeneracy. It is worth remarking on another prominent consequence regarding the new dual structure and its definition. Differently from Dirac spinors, the whole class of singular spinors are not eigenstates of the parity operator \cite{rrmpla21}. Regarding Elko spinors,  the most important singular spinor representative for DM physics, it obeys the coupled relations below\footnote{ Here, $\slashed{p}\equiv \gamma^\mu p_\mu$ is proportional to the parity operator $\frac{\slashed{p}}{m}$.}
\bea \slashed{p}\uplambda^S_{\{+,-\}}(\boldsymbol{p})=\pm im \uplambda^S_{\{-,+\}}(\boldsymbol{p}), \qquad \qquad \slashed{p}\uplambda^A_{\{+,-\}}(\boldsymbol{p})=\mp im \uplambda^A_{\{-,+\}}(\boldsymbol{p}) \label{Elkop},\eea
laddering between the Elko types present in the definition of the previously mentioned new basis, introduced in compliance with the Wigner degeneracy. These equations ensure that all Elko spinors obey the Klein-Gordon equation, a necessary condition to provide a fundamental particle description. 

Another relevant physical definition for this work lies in the structure of the quantum field. Once we have introduced the complete set of $\xi$ and $\varrho$ singular spinors, we can move towards their definition\footnote{For more details, please, check Ref. \cite{rodolfosingular}.}
\begin{eqnarray}\label{campoquanticofinal}
\!\!\!\!\!\!\!\!\!\uplambda(x) =\int\frac{d^3 p}{(2\pi)^3}
\frac{1}{\sqrt{ m(|\alpha|^2+|\beta|^2) E(\p)}}
\bigg[
\sum_{h} {c_h}(\p)\xi_h(\p) e^{-i p\cdot x}+ \sum_{h} d^\dagger_h(\p)\upchi_h(\p) e^{i p\cdot x}\bigg],
\end{eqnarray}
and the associated dual 
\begin{eqnarray}\label{campoquanticodualfinal}
\!\!\!\!\!\!\!\!\!\gdualn{\uplambda}(x) = \int\frac{d^3 p}{(2\pi)^3}
\frac{1}{\sqrt{ m(|\alpha|^2+|\beta|^2) E(\p)}}
\bigg[
\sum_{h} c^\dagger_h(\p)\gdualn{\xi_h}(\p) e^{i p\cdot x}+ \sum_{h} d_h(\p)\gdualn{\upchi_h}(\p)e^{-i p\cdot x}\bigg],
\end{eqnarray} 
with ${c}^\dagger_h(\p)$ and ${d}^\dagger_h(\p)$ denoting the particle and anti-particle creation operators, respectively.\\
\indent This field expansion defines a mass dimension one fermion whose Lagrangian for free field reads \bea \mathcal{L}=\partial_\mu \gdualn{\uplambda}(x)\partial^\mu \uplambda(x)-m^2 \gdualn{\uplambda}(x) \uplambda(x),       \eea encoding the fact that the only uncoupled equation fulfilled by this field is the Klein-Gordon one, defining a relativistic particle. The mass dimension one nature avoids the inclusion of such spinors in renormalizable interactions in the form of standard model-like doublets, being another signature compatible with DM. It is worth mentioning that the quantum field whose expansion coefficients are the Elko spinors ($\alpha=i$ and $\beta=1$)  defines intrinsically neutral particles whereas the field itself is a combination of self and (anti) self-conjugate spinors under charge conjugate operator. These properties evince the special role played by Elko spinors in establishing a DM field. 

The aforementioned properties yield the following  Feynman-Dyson propagator for the whole class of singular spinors
\begin{align}\label{propagadorfinal}
S_{\textrm{FD}}(x^\prime-x)=\int\frac{\text{d}^4 p}{(2 \pi)^4}\,
e^{-i p_\mu(x^{\prime\mu}-x^\mu)}
\frac{\mathbb{1}}{p_\mu p^\mu -m^2 +i\epsilon},
\end{align} 
implying a local and Lorentz invariant structure. It also leads to a well-defined Hamiltonian compatible with a standard particle interpretation.  Although the model's differential operator puts no restriction on the spinor structure according to Ref. \cite{Ahluwalia:2022ttu}, a consistent rotationally invariant QFT  for DM in terms of external Elko particles can be achieved only through the consideration of the correct dual structure in the context of Wigner degeneracy, characterizing this specific system. Namely, the enlarged basis leads to four particles and four anti-particles. On the other hand, since the theory has no constraints, a Hamiltonian analysis reveals sixteen phase space degrees of freedom. Fortunately, it precisely corresponds to the eight configuration space excitations comprised by the free field construction based on Wigner's degeneracy. This correspondence addresses the correct approach to formulating the theory.

\section{On action formulation, interactions, and the Hermitian nature} \label{3}

\indent From now on, in order to model dark matter, just the quantum field based on the Elko spinor expansion coefficients will be regarded. In this case, the spinor expansion coefficients of the particle and anti-particle sectors are, respectively, self-conjugate and anti-self-conjugate under the action of the charge conjugation operator.  The free field structure is given by
\bea \uplambda(x)=\int \frac{d^3p}{(2\pi)^3\sqrt{2mE(p)}}\left[ \left(\sum_{h=1}^{4}c_h(\p)\xi_h(\p)  \right)e^{-ip.x}+\left(\sum_{h=1}^{4}d^\dagger_h(\p)\upchi_h(\vec p)  \right)e^{ip.x}   \right],            \eea
whose dual reads \footnote{Using the definition $\mathcal{P}=\frac{\slashed{p} }{m}$.}
\bea \gdualn{\uplambda}(x)=\int \frac{d^3p}{(2\pi)^3\sqrt{2mE(p)}}\left[ \left(\sum_{h=1}^{4}c^\dagger_h(\p)(\mathcal{P}\xi_h(\vec p))^\dagger  \right)e^{ip.x}+\left(\sum_{h=1}^{4}d_h(\p)(-\mathcal{P}\upchi_h(\p))^\dagger  \right)e^{-ip.x}   \right] \gamma_0  .         \eea
As an important observation, considering the structure of the parity operator, the dual field can be expressed as 
\bea\label{gdd} \gdualn{\uplambda}(x)=\left(i\frac{\slashed{\partial}}{m} \uplambda(x)\right)^\dagger \gamma_0.             \eea
 Considering Eq. (\ref{gdd}) one notices that the action below is Hermitian, up to a boundary term, 
\bea \mathcal{S}=\int d^4x \left[\partial_\mu \gdualn{\uplambda}(x)\partial^\mu\uplambda(x)-m^2\gdualn\uplambda(x)\uplambda(x)              \right].\label{lagr} \eea
Therefore, rewriting the scalar bilinear as  \bea \gdualn{\uplambda}(x)\uplambda(x)=-i\uplambda^\dagger(x)\gamma_0 \Big(\frac{\gamma^\mu \stackrel{\leftarrow}{\partial_\mu}}{m}\Big)^{ \dagger}\uplambda(x),    \eea
  using the identity $\frac{\slashed{\partial}}{m}\gamma_0=\gamma_0\frac{\slashed{\partial}}{m}$, and discarding a total derivative term, it immediately leads to
\bea \mathcal{S}=\mathcal{S}^\dagger. \eea

A Hermitian and renormalizable interaction with the Higgs boson can be described by the following Lagrangian term
\bea \mathcal{L}_I=\gdualn{\uplambda}(x)i\slashed{\partial}\uplambda(x)\phi(x)g. \eea 
It is straightforward to show that the Lagrangian itself is fully Hermitian, since
\begin{eqnarray}
    \Big[\frac{1}{m}(i\slashed{\partial}\uplambda(x))^\dagger\gamma_0(i\slashed{\partial}\uplambda(x))\phi(x) g    \Big]^\dagger&=&\Big[\frac{1}{m}(i\slashed{\partial}\uplambda(x))^\dagger\gamma_0^\dagger \Big((i\slashed{\partial}\uplambda(x))^\dagger\Big)^\dagger \phi(x) g    \Big]\nonumber\\&=&\gdualn{\uplambda}(x)i\slashed{\partial}\uplambda(x)\phi(x)g, \end{eqnarray} 
    where $g$ is a dimensionless coupling.\\
    \indent At this point it is worth mentioning the fact that the quadratic part of the action is invariant under the charge conjugation symmetry, whose operation in the mass dimension one field reads $\uplambda^c(x)=\gamma_2\uplambda^*(x)$, see Refs. \cite{dharamnpb,Ahluwalia:2022ttu}. Accordingly, the Elko-Higgs interaction is also invariant under such a discrete symmetry transformation already at the Lagrangian level.\\
    \indent Regarding the free model, the U$(1)$ fermion number symmetry implies the following conserved current
    \bea  J_\mu(x)=i\Big( \gdualn{\uplambda}(x)\partial_\mu \uplambda(x)-\partial_\mu \gdualn{\uplambda}(x) \uplambda(x)     \Big),\eea
    whose associated charge $Q=\int d^3x J_0(x)$ indeed changes sign under the operation of charge conjugation, up to non-contributing boundary terms. However, a minimal interaction with electromagnetism is based on a linear coupling with such current functional as well as a non-derivative quartic term. Then, the theory becomes not invariant under charge conjugation even if the electromagnetic field changes its sign, since the trick based on discarding surface terms cannot be done for the trilinear field interaction term. Moreover, the resulting theory would be non-Hermitian, a second strong reason that avoids standard minimal couplings between mass dimension one fields and the electromagnetic potential. Thus, the features mentioned above define the dark nature of such a spinor field.

\section{Hermitian interactions and a criterion for a dark coupling}\label{5}
\indent The previous constructions allow us to derive a set of Hermitian interactions suitable to, at least, define effective models. The first one consists of a 
Yukawa-like interaction with the Higgs boson, given by 
\bea \mathcal{L}_I=g\gdualn{\uplambda}(x)i\slashed{\partial}\uplambda(x)\phi(x).  \eea 
It furnishes amplitudes whose structures are different from all the known ones from the standard model. It is renormalizable and Hermitian.\\
\indent One can also define a Hermitian extension of the neutral Pauli-like coupling with the electromagnetic field, as
\bea  \mathcal{L}_I=g'\gdualn{\uplambda}(x)i\left[\gamma^\mu,\gamma^\nu \right]F_{\mu \nu}(i\slashed{\partial}\uplambda(x)),                 \eea
with $g'$ being a coupling with a dimension of inverse powers of mass.
It is worth mentioning that the previous pseudo-Hermitian version is associated with a darkness mechanism regarding the coupling with mass dimension one fields constructed from Elko spinors  \cite{NPB2023}.

The Elko effective coupling with Dirac fermions such as the neutrino,  explored in 
 a phenomenological context  in Ref. \cite{Moura:2021rmf}, can be deformed to become Hermitian as
\bea  \mathcal{L}_I=\mathrm{c}_1\gdualn{\uplambda}(x)(i\slashed{\partial})\uplambda(x)\Bar \eta(x) \eta(x),                 \eea
with $\mathrm{c}_1$ being another dimensionful coupling parameter as well as $\mathrm{c}_2$ and $\mathrm{c}_3$ present in the next two interactions highlighted in this section.

The  Elko-Higgs coupling with four legs is Hermitianized as
\bea   \mathcal{L}_I=\mathrm{c}_2\gdualn{\uplambda}(x)(i\slashed{\partial})\uplambda(x)\phi^2(x),                \eea 
 whereas the  self-interaction can have a Hermitian version like
\bea    \mathcal{L}_I=\mathrm{c}_3\Big(\gdualn{\uplambda}(x)(i\slashed{\partial})\uplambda(x) \Big)^2.          
   \eea
\noindent Therefore, although these approaches define legitimate models by themselves, we consider that the final theory must undergo renormalizability and Hermiticity, defining our physical requirements. This criterion fixes the Yukawa-like coupling as a viable portal for investigating DM physics. As we will see, this interaction is suitable to explain some of the observed experimental data in a natural setting.\\
\indent Last but not least, one can mention the possibility of the so-called pseudo-Hermitian interactions, in connection with previous works. They are associated with a quartic coupling with Higgs and a self-interaction for the mass dimension one fields \cite{Ahluwalia:2022ttu}. This approach was also successfully adopted for the gravitational interaction of fields based on Elko spinors, in Ref.  \cite{Lee:2024sbg}. A suitable procedure to extract a well-defined probabilistic interpretation for these couplings is currently under development, see the latest achievements in Ref. \cite{Ahluwalia:2023slc}. It can lead to a wider possibility of interactions, in addition to the Hermitian ones highlighted above, improving the modeling power. 

\section{Elko-Higgs derivative coupling}\label{6}

\begin{figure}[H]
\centering
\includegraphics[width=13.15cm] {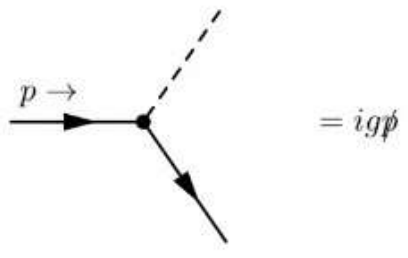}  \vspace*{-7.5cm}
\caption{Elko-Higgs coupling. The arrows denote mass dimension one fields and the dashed line refers to the Higgs particle.}
\label{full}
\end{figure}

\indent Taking into account Hermiticity and renormalizability, we consider the derivative Yukawa theory as the main paradigm. Hence, the following action is considered
\bea \!\!\!\!\!\!\!\!\!\!  S\!=\!\!\int\! d^4x\Bigg[\partial_\mu \gdualn{\uplambda}(x)\partial^\mu\uplambda(x)\!-\!m^2\gdualn\uplambda(x)\uplambda(x)\!+\!\frac{1}{2}\big(\partial_\mu \phi \partial^\mu \phi\!-\!M^2\phi^2\big) \!+\!\gdualn{\uplambda}(x)i\slashed{\partial}\uplambda(x)\phi(x)g\!+\!\cdots  \Bigg], \eea
where $\cdots$ designates all the remaining Higgs couplings in the electroweak sector as well as its quartic self-interaction.\\
\indent The interaction is associated with the vertex shown in Fig. \ref{full}.  
%\begin{fmffile}{fp}
%\begin{gathered}
 %   \quad\parbox{100pt}{
  %  \begin{fmfgraph*}(80,80)
   %    \fmfleft{i}
    %   \fmfright{o1,o2}
     %  \fmfv{label=$ p \to$,l.a=60}{i}
       %\fmfv{label=p,l.a=120}{o}
     %  \fmf{fermion,tension=1}{i,v1} % ,label=H,label.side=left
       %\fmf{dashes,tension=1}{v2,o}
       %\fmf{photon,label=$\text{k}$,tension=0.4}{v1,o3}
     %  \fmf{photon,tension=1}{v1,o2}
     %  \fmf{fermion,tension=1}{v1,o1}
      % \fmfdot{v1}
     %  \end{fmfgraph*}}
     %  \end{gathered}=ig\slashed{p}
     % \end{fmffile}
    \vspace*{-8cm}
      
\begin{figure}[H]
\centering
\includegraphics[width=14cm]{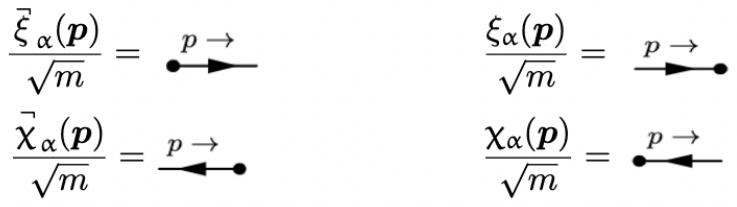} \vspace*{-8.5cm}
\caption{Feynman rules for external Elko fields.}
\label{full2}
\end{figure}

       The Feynman rules for external Elko fields are depicted in Fig. \ref{full2}. \footnote{The $h$ label is replaced by the $\alpha$ one. This is just a notational change convenient for the next discussions.}
%\frac{\gdualn{\xi}_\upalpha(\p)}{\sqrt{m}}
   % =  %\vcenter{\hbox{\includegraphics[width=1.4cm,height=0.85cm]{ext1.png}}}
    %\qquad\qquad\qquad\frac{\xi_\upalpha(\p)}{\sqrt{m}}
    %=  \vcenter{\hbox{\includegraphics[width=1.4cm,height=0.85cm]{ext2.png}}}\nn\\
    %\frac{\gdualn{\upchi}_\upalpha(\p)}{\sqrt{m}}
    %= \vcenter{\hbox{\includegraphics[width=1.35cm,height=0.9cm]{ext3}}}\qquad\qquad\qquad
      %\frac{\upchi_\upalpha(\p)}{\sqrt{m}}
   % =  \vcenter{\hbox{\includegraphics[width=1.35cm,height=0.9cm]{ext4}}}
   \vspace*{-8.5cm}

The mass dimension one field associated with DM has the following propagator 
\bea \frac{i}{k^2-m^2+i\epsilon},\eea
whereas the Higgs propagator reads
\bea  \frac{i}{k^2-M^2+i\epsilon}.\eea
\noindent Therefore, from these building blocks, one can derive the 1-loop radiative corrections and explicitly discuss renormalizability. Later, a set of scattering processes can be analyzed,  to define the role of mass dimension one fields constructed in terms of Elko spinors in the context of the DM phenomenology. The Feynman rules also reveal another important aspect of theories based on the Elko spinors: Although the mass dimension one quantum field is not invariant under charge conjugation, the external states define its eigenstates. Moreover, we verified that all the amplitudes studied here based on the  Elko-Higgs interaction, which is charge conjugation invariant, are such that $\mathcal{M}^*=\mathcal{M}^c$, according to the anti-unitary nature of this discrete symmetry operation. Here, $\mathcal{M}^c$ denotes a given amplitude in which the external states are replaced by their charge-conjugated versions. Since the processes related by this discrete symmetry necessarily keep the external Elko sector, it can be understood as another darkness  signature.  
\section{Renormalizability at one loop}\label{7}

This section is devoted to discussing renormalizability. We explicitly compute the divergent $n$-point functions at 1-loop approximation and show that the singular pieces have the same form as the terms originally present in the bare Lagrangian. It means that they can be absorbed in a consistent renormalization scheme such as the on-shell one, for example. Although power-counting arguments play an important role in quantum field theory, there are cases in which even a power-counting renormalizable theory may need an ultraviolet completion to ensure finiteness. One can mention the illustrative case of the quantum scalar electrodynamics \cite{rohr}, in which it is necessary to add an extra self-interaction term for the scalar particles, despite the good power-counting arguments for this specific interaction. Therefore, a careful explicit evaluation of the 1-loop renormalized structure is required.\\
\indent Considering the Feynman parametrization, the bosonic self-energy reads 
\bea\label{divvf}
 i\Pi_\phi(p)= g^2\int \frac{d^4k}{(2\pi)^4}\frac{Tr[(\slashed{p}-\slashed{k})\slashed{k}]}{[(p\!-\!k)^2\!-\!m^2][p^2\!-\!m^2]}=g^2\!\!\int \!\frac{d^4k}{(2\pi)^4}\int_0^1\! dx\frac{Tr[p^2x(1\!-\!x)\!-\!k^2]}{[k^2-\Delta_x]^2},\eea
with $\Delta_x=m^2-p^2x(1-x)$. The divergent part of Eq. \eqref{divvf} wits \footnote{Considering dimensional regularization in the limit $\Bar \epsilon \to 0$. }
\bea \left(\Pi_\phi(p)\right)_{div}= \frac{g^2}{16\pi^2\Bar \epsilon}\,p^2-m^2 \frac{g^2}{8\pi^2\Bar \epsilon},         \eea
which has indeed the same type of the bare Lagrangian.\\
\indent The fermionic self-energy is given by
\bea i\Sigma(p)=g^2\int \frac{d^4k}{(2\pi)^4}\frac{\slashed{k}\slashed{p}}{(k^2-m^2)\left((k-p)^2-M^2\right)}=p^2\int \int_0^1dx  \frac{d^4k}{(2\pi)^4}\frac{x}{\left(k^2-\Delta'_x\right)^2},                      \eea for 
$\Delta'_x=(1-x)(m^2-p^2x)+xM^2$. The divergent piece has the same form as the  Gaussian kinetic term in the bare Lagrangian and reads
\bea    \Big(\Sigma(p)\Big)_{div}=\frac{g^2}{16\pi^2\Bar \epsilon}\,p^2,   \eea
ensuring renormalizability. 
Hence, the remaining divergent radiative correction is the vertex function
\bea  i\Gamma(q_1,p)= g^3\int \frac{d^4k}{(2\pi)^4}\frac{[(\slashed{p}+\slashed{k}) \slashed{k} \slashed{q}_1]}{[(k+q_1)^2-M^2][k^2-m^2 ][(p+k)^2-m^2 ]}, \eea
with $q_1$ being the fermionic momentum entering the graph in the same arrow orientation. It can be parametrized as
\bea     [i\Gamma(q_1,p)]_{div}=\slashed{q}_1 g^3\int \int dx\,dy\,dz \,\delta(x+y+z-1)\frac{d^4k}{(2\pi)^4}\frac{k^2}{[k^2-\tilde{\Delta}_x]^3}, \eea                              with $\tilde{\Delta}_x=-xyp^2+(1-z)^2m^2$. Its divergent piece has also the same form as bare Lagrangian
\bea     [\Gamma(q_1,p)]_{div}=\slashed{q}_1 \frac{g^3}{16\pi^2  \Bar \epsilon},  \eea  
completing the leading-order renormalizability verification.\\
\indent According to the next developments, it is possible to define the model's parameter configurations in compliance with experimental data in which $g$ is not prohibitively small, meaning that the study of the radiative corrections is relevant for this phase.\\
\indent Then, summarizing, the presence of the derivative interaction adds a power of the integration momentum in each fermion internal line. Moreover, considering the specific structure of the DM propagator and the fact that the trilinear graph topology is the same as the standard QED$_4$ one, it is possible to show that the superficial degree of divergence resembles the expression concerning QED$_4$ \footnote{Here, $N_\phi$ and $N_\uplambda$ denote the number of external Higgs fields and the number of external Elko, respectively.},
\bea D=4-N_\phi-\frac{3}{2}N_\uplambda,    \eea
meaning that the four-point 1PI function with external bosonic lines is logarithmic-divergent. Differently from  QED case, there is no symmetry implying the decrement of the divergence degree for this specific radiative correction. Fortunately, the divergent piece is constant and can be absorbed in a renormalization procedure involving the bare Higgs four legs vertex, completing our 1-loop analysis.

\section{M\o ller-like scattering}\label{8}

\indent The Hermitian interaction with the Higgs boson leads to an interesting phenomenology due to the wide set of electroweak couplings involving the Higgs. In order to begin this discussion, one can evaluate the M\o ller-like scattering between Elko particles. The appropriate scattering parametrization in the center-of-mass frame reads
\begin{subequations}
\bea    p_1^\mu&=&(E,0,0,p), \qquad\qquad\qquad\qquad p_2^\mu=(E,0,0,-p),\\
p_3^\mu&=&(E, p \sin \theta,0, p \cos\theta), \qquad\qquad p_4^\mu=(E, -p \sin \theta, 0, -p \cos\theta),              \eea
\end{subequations}
with $E=\sqrt{p^2+m^2}$. This scattering process has contributions from the $t$ and $u$ channels with the following associated amplitude
\bea
    i{\cal{M}}_{\upalpha, \upbeta, \upbeta^\prime, \upalpha^\prime}&=&\frac{g^2}{m^2\big (t-M^2\big)}i\gdualn  
 \xi_\upalpha(\p_4)\slashed{p}_2\xi_\upbeta(\p_2) \gdualn \xi_{\upbeta'}(\p_3)\slashed{p}_1\xi_{\upalpha'}(\p_1)\nonumber \\
    &&-\frac{g^2}{m^2 \big(u-M^2\big)}i\gdualn \xi_{\upbeta'}(\p_3)\slashed{p}_2\xi_{\upbeta}(\p_2) \gdualn \xi_{\upalpha}(\p_4)\slashed{p}_1\xi_{\upalpha'}(\p_1),
\eea
whereas the Hermitian conjugate reads
\beq
    \!\!\!\!\!\!\!\!\!\!{\cal{M}}^\dagger_{\upalpha, \upbeta, \upbeta^\prime, \upalpha^\prime}&=&\frac{g^2}{m^2\big(t-M^2\big)}\gdualn \xi_{\upalpha'}(\p_1)\slashed{p}_3\xi_{\upbeta'}(\p_3)\gdualn \xi_{\upbeta}(\p_2)\slashed{p}_4\xi_\upalpha (\p_4)\nn \\
    &&-\frac{g^2}{m^2\big(u-M^2\big)}\gdualn \xi_{\upalpha'}(\p_1)\slashed{p}_4\xi_{\upalpha}(\p_4)\gdualn \xi_\upbeta(\p_2)\slashed{p}_3\xi_{\upbeta'}(\p_3). 
\eeq
\indent Therefore, the non-polarized squared amplitude reads \footnote{Use has been made of the identities $Tr(\gamma^\mu \gamma^\nu \gamma^\alpha \gamma^\beta)=Tr(\gamma^\nu \gamma^\mu \gamma^\beta \gamma^\alpha)$ and $Tr(\gamma^\mu \gamma^\nu \gamma^\alpha \gamma^\beta)=Tr(\gamma^\mu \gamma^\beta \gamma^\alpha \gamma^\nu)$. }
\beq
    \!\!\!\!\!\!\!\!\!\!\frac{1}{16}\sum_{\upalpha,\upalpha^\prime,\upbeta,\upbeta^\prime}{\cal{M}}_{\upalpha, \upbeta, \upbeta^\prime, \upalpha^\prime}{\cal{M}}^\dagger_{\upalpha, \upbeta,\upbeta^\prime,\upalpha^\prime}=\frac{1}{16}\Bigg[&A^2& Tr(\slashed{p}_1\slashed{p}_3)Tr(\slashed{p}_2\slashed{p}_4)+B^2Tr(\slashed{p}_1\slashed{p}_4)Tr(\slashed{p}_2\slashed{p}_3) \nonumber \\&-&2AB\ Tr(\slashed{p}_4 \slashed{p}_2  \slashed{p}_3 \slashed{p}_1  )        \Bigg].
\eeq
Here, we considered the following label sums $\sum_\alpha \xi_\alpha(\p)\gdualn{\xi}_\alpha(\p)=m\boldsymbol{1}=-\sum_\alpha \chi_\alpha(\p)\gdualn{\chi}_\alpha(\p)$ as well as the definitions $A=\frac{g^2}{(t-M^2)}$ and $B=\frac{g^2}{(u-M^2)}$.\\
\indent Regarding the squared polarized amplitudes, we can mention some of them like \footnote{They are explicit positive definite, complying with the Hermitian nature.} 
\begin{eqnarray}\label{ampli111}
 |{\cal{M}}_{1,1,1,1}|^2 = \left(
\begin{array}{c}
 \frac{g^4  p^4 \sin ^4(\theta ) (m+2 E )^4 (p^2-E^2)^2 (t-u)^2 \left((m+E)^2-p^2\right)^2 \left(-M^2+t+u\right)^2}{16 m^{12}  (m+E )^4 \left(M^2-t\right)^2 \left(M^2-u\right)^2} \\
\end{array}
\right),   
\end{eqnarray}
highlighting the fact that the very same result is obtained for the cases $|{\cal{M}}_{2,2,2,2}|^2$, $|{\cal{M}}_{3,3,3,3}|^2$, and $|{\cal{M}}_{4,4,4,4}|^2$.

One can also mention other kinds of transitions such as
\begin{eqnarray}\label{ampli1122}
 &&|{\cal{M}}_{1,1,2,2}|^2 =\left(
\begin{array}{c}
 \frac{g^4 p^4 \sin ^4(\theta ) (m+2 E )^4 (p+E )^4 (t-u)^2 (m+p+E )^4 \left(-M^2+t+u\right)^2}{16 m^{12}   (m+E )^4 \left(M^2-t\right)^2 \left(M^2-u\right)^2} \\
\end{array}
\right),
\end{eqnarray}
and 
\begin{eqnarray}\label{ampli3434}
  |{\cal{M}}_{3,4,3,4}|^2 =  
\begin{array}{c}
% \frac{g^4  (p^2-E^2 )^2  (t-u)^2 \big((m+E)^2-p^2 \big)^2  \left(-M^2+t+u\right)^2 \left(-p \cos (\theta ) (m+2 E )+m (m+E )+2 p^2\right)^2 \left(p \cos (\theta ) (m+2 E)+m (m+E )+2 p^2\right)^2}{16 m^{12}  (m+E )^4 \left(M^2-t\right)^2 \left(M^2-u\right)^2} \\
 \frac{g^4  (p^2-E^2 )^2  (t-u)^2 \big((m+E)^2-p^2 \big)^2  \left(-M^2+t+u\right)^2 \left[-\left(p \cos (\theta ) (m+2 E )\right)^2+\left(m (m+E )+2 p^2\right)^2\right]^2}{16 m^{12}  (m+E )^4 \left(M^2-t\right)^2 \left(M^2-u\right)^2} \\
\end{array}.
\end{eqnarray}
Interestingly, in Appendix \ref{a1} we generalize this result for the whole set of singular spinors and comment on possible relevant special cases.

\section{On the Elko annihilation processes}\label{9}

\indent The study of Elko annihilation processes is a useful tool to constrain the model's parameters through cosmological requirements associated with the current measurements of DM abundance. As we will see,  the coupling with the Higgs ensures a rich DM phenomenology that complies with experimental data. \\
\indent Regarding the Elko annihilation processes, the situation is suitably parameterized by the variables
\begin{subequations}
\bea    p_1^\mu&=&(E, 0, 0, p), \qquad\qquad\qquad\qquad p_2^\mu=(E, 0, 0, -p),\\
p_3^\mu&=&(E, p' \sin \theta, 0, p' \cos\theta), \qquad\qquad p_4^\mu=(E, -p' \sin \theta, 0, -p' \cos\theta),              \eea
\end{subequations}
with $p_1^\mu$ and $p_2^\mu $ representing the 4-momentum of the incoming DM particles in the center-of-mass frame. The remaining variables denote the 4-momentum of the outgoing produced particles. For the case of an annihilation process of two incoming Elko into two outgoing Higgs particles, one can set the energy as $E=\sqrt{p^2+m^2}=\sqrt{p^{'2}+M^2}$, with $m$ and $M\approx 125$ GeV representing the Elko spinor and Higgs masses, respectively. \\
\indent Considering the Feynman rules, the amplitude reads
\bea i{\cal{M}}_{\beta \sigma}=i\frac{g^2}{m(t-m^2)} \gdualn{\chi}_\beta(\p_2)(\slashed{p}_1-\slashed{p}_3)\slashed{p}_1 \xi_\sigma(\p_1)+i\frac{g^2}{m(u-m^2)}\gdualn{\chi}_\beta(\p_2)(\slashed{p}_1-\slashed{p}_4)\slashed{p}_1 \xi_\sigma(\p_1).                     \eea
\indent For cosmological considerations, the relevant quantity is the non-polarized amplitude \footnote{Using the identity $(\gdualn{\chi}_\alpha(\p_2) \slashed{p}_1 \xi_\delta(\p_1))^\dagger=-\gdualn{\xi}_\delta(\p_1)\slashed{p}_2\chi_\alpha(\p_2)$.}
\bea \frac{1}{4}\sum_{\beta \sigma} |{\cal{M}}_{\beta \sigma}|^2=\frac{C^2}{4}Tr\left[((\slashed{p}_1-\slashed{p}_3))\slashed{p}_1(\slashed{p}_1-\slashed{p}_3)\slashed{p}_2\right]+\frac{D^2}{4}Tr\left[((\slashed{p}_1-\slashed{p}_4))\slashed{p}_1(\slashed{p}_1-\slashed{p}_4)\slashed{p}_2\right] \nonumber \\   +\frac{CD}{4}Tr\left[((\slashed{p}_1-\slashed{p}_4))\slashed{p}_1(\slashed{p}_1-\slashed{p}_3)\slashed{p}_2 \right] +\frac{CD}{4}Tr\left[((\slashed{p}_1-\slashed{p}_4))\slashed{p}_2(\slashed{p}_1-\slashed{p}_3)\slashed{p}_1 \right]  \eea
with the definition  $C=\frac{g^2}{(t-m^2)}$ and $D=\frac{g^2}{(u-m^2)}$, in terms of the $t$ and $u$ channel Mandelstam variables.\\ 
\indent Since the Higgs boson presents  Yukawa-like couplings with a variety of particles, one should also consider $s$-channel annihilation processes into a pair of fermions due to the electroweak couplings of the form $g_{\mathrm{J}}\phi(x) \Bar\psi_{\mathrm{J}}(x) \psi_{\mathrm{J}}(x)$, in which the label ${\mathrm{J}}$ regards the fermionic types ${\mathrm{J}}=e,u,d$ respectively associated with the electron, the up, and the down quarks. The amplitude of the process reads \footnote{Here, $\Bar u_{\gamma'}\mathrm{J}(p_3)$ and $v_{\sigma'}^\mathrm{J}(p_4)$ denote the electron and positron spinors, respectively. The label $\gamma$ and $\sigma$ are associated with the spin.}
\bea i{\cal{M}}_{\alpha \beta \gamma' \sigma' }=-i\frac{gg_{\mathrm{J}}}{m(s-M^2)} \gdualn{\chi}_\alpha(\p_1)\slashed{p}_2 \xi_\beta(\p_2) \Bar u_{\gamma'}^{\mathrm{J}}(p_3) v_{\sigma'}^{\mathrm{J}}(p_4).                   \eea
\indent The non-polarized squared amplitudes has the following expression
\bea \frac{1}{8}\sum_{\alpha \beta \gamma' \sigma' }|{\cal{M}}_{\alpha \beta \gamma' \sigma' }|^2=\frac{(gg_{\mathrm{J}})^2}{8(sM^2)^2}Tr\left[\slashed{p}_1\slashed{p}_2\right]Tr\Big[(\slashed{p}_3+m_\mathrm{J})(\slashed{p}_4-m_\mathrm{J})\Big],            \eea
with $m_e\approx 0.5$ MeV,  $m_u\approx 2.15$ MeV and $m_d\approx 4.5$ MeV. The coupling constants have magnitude $g_e\approx 2\times 10^{-6}$, $g_u\approx  10^{-5}$,  and $g_d\approx 2\times 10^{-5}$. It is worth mentioning that according to further developments in the next sections, this weak coupling is the origin of the difficulty in observing DM/electron scattering in the laboratory, which means that the investigation of this portal leads to a natural description of such phenomenology.\\
\indent Beyond the $s$-channel annihilation into fermion pairs, one can also consider this kind of annihilation involving vector bosons. They are associated with the electroweak Higgs couplings of the form \footnote{The symbols $V^\mu_A(x)$ denote the vector fields representing the $Z$ and $W$ vector bosons. }
\bea g_A m_A\phi(x)\eta_{\mu \nu}|V^\mu_A(x)| |V^\nu_A(x)|,\eea with $A=Z,W_+,W_-$ recovering the cases of the weak $Z$ and $W$ vector bosons, respectively. The amplitude reads \footnote{Here $\epsilon_\mu^r$ represent the polarization vectors of the vector boson fields.}
\bea i{\cal{M}}_{\alpha, \beta r,s}=-im_A\frac{gg_A}{m(s-M^2)} \gdualn{\chi}_\alpha(\p_1)\slashed{p}_2 \xi_\beta(\p_2)\eta^{\mu \nu}\epsilon_\mu^r\epsilon_\nu^s.                    \eea
The averaged non-polarized amplitude for these massive bosons
\footnote{ Considering the relation $\sum_r\epsilon_\mu^r \epsilon_\nu^{*r}=-\eta_{\mu \nu}+\frac{p_\mu p_\nu}{m^2_A}$ for the polarization vectors.}
\bea \frac{1}{12}\sum_{\alpha \beta r s }|{\cal{M}}_{\alpha \beta r s }|^2=\frac{m^2_A(gg_A)^2}{4(s-M^2)^2}Tr\left[\slashed{p}_1\slashed{p}_2\right],     \eea
with $m_Z\approx 91.19$ GeV and $m_W\approx 80.38$ GeV, whereas for the  couplings one has $g_z\approx 0.36$ and $g_w\approx 0.32 $.\\
\indent Therefore, after deriving the amplitudes, the cross-section defined as 
\bea \sigma=\int \frac{1}{64\pi^2E_{cm}^2}\frac{|\vec {p}_f||\mathcal{M}|^2\ d\Omega}{|\vec{p}_i|}    \eea
is another pivotal quantity that enters the Boltzmann equation governing the dynamics of the Elko density in the presence of such annihilation processes.

\section{Investigating the Freeze-out process}\label{10}
\indent The Boltzmann equation describing the dynamics of the DM density $n$ in the presence of $2\to 2 $ scattering processes in an expanding universe, reads \cite{Bauer:2017qwy}
\bea   \label{dmd} \frac{d n}{dt}+3Hn=\langle \sigma v\rangle \Big(n^2-n^2_{eq}\Big) \eea
with $H$ being the Hubble constant and $n_{eq}$ denoting the equilibrium density.\\
\indent As the universe expands, it begins to cool down, implying lower rates for annihilation processes. When it reaches $\Gamma(T_{eq}) \approx H$ the density of DM stabilizes in its relic value, for $T_{eq}$ denoting the freeze-out temperature. Here, $\Gamma$ is the annihilation rate in the thermal average. In a cold DM scenario, it occurs at a temperature $T< m$ with a non-relativistic distribution $n\sim g^* \Big(\frac{mT}{2\pi}\Big)^{3/2}e^{-m/T}$, with $g^*$ being the degrees of freedom associated to the reaction. For this kind of $2\to 2$ processes with DM initial states, one can set $g^*=8$ considering the Elko field in compliance with the Wigner degeneracy structure. In non-relativistic limit $\Gamma= \langle \sigma v\rangle n \sim \sigma v n$, with the relative velocity of the incoming particles being temperature-dependent as  $v\approx 2\sqrt{\frac{2T}{m}}$. The cold DM scenario is supported by the cosmological model $\Lambda$CDM which is the one with the sharpest agreement with observations of the CMB in the context of an expanding flat universe \cite{Planck:2018nkj,WMAP:2010qai}. \\
\indent The Boltzmann equation can be managed to yield a constraint involving the thermal averaged amplitude and the cosmological DM  abundance 
\bea \label{dmd1}
%\Omega_{DM}h^2=0.12 \frac{x_{eq}}{23}\frac{\sqrt{g_{eff}}}{10}\frac{1.7\times 10^{-9}GeV^{-2}}{\langle \sigma v\rangle }
 \Omega_{DM}h^2=8.869\times 10^{-13}\,{\rm GeV}^{-2}\,\frac{x_{eq}\sqrt{g_{eff}}}{\langle \sigma v\rangle},
\label{abund} \eea
 with $x_{eq}\equiv \frac{m}{T_{eq}}$. For the freeze-out of a non-relativistic species $x_{eq}$ is typically of order $\approx 24$ for DM with mass $\approx 100$ GeV. The observed cosmological phenomenology fixes the abundance of DM as $\Omega_{DM}h^2\approx 0.12$. Here, since a weak scale mass is considered in a cold DM scenario, the expected freeze-out temperature is such that the effective thermal degrees of freedom $g_{eff}\approx 86.25$, see Ref. \cite{Bauer:2017qwy}.\\
 \indent The amplitude squared for the Elko pair annihilation into Higgs bosons in the low energy limit is \footnote{ $T\ll m$, $p^2=2mT$. We also consider that $t$-channel is irrelevant compared to $m$ in denominator since a non-relativistic freeze out is being regarded.}
 \bea \frac{1}{8}\sum_{\alpha \beta \gamma' \sigma' }|{\cal{M}}_{\alpha \beta \gamma' \sigma' }|^2\approx \frac{8g^4T}{m}.            \eea
\indent The non-polarized squared amplitudes for Elko pair annihilation into fermionic particle and anti-particle pairs are well approximated by
 \bea  \frac{1}{8}\sum_{\alpha \beta \gamma' \sigma' }|{\cal{M}}_{\alpha \beta \gamma' \sigma' }|^2\approx 8\frac{(gg_{\mathrm{J}})^2m^2m_{\mathrm{J}} T}{(2m^2+2m^2_{\mathrm{J}}-M^2)^2}, \eea
 whereas the annihilation into vector bosons are associated with the limiting value
 \bea  \frac{1}{12}\sum_{\alpha \beta r s }|{\cal{M}}_{\alpha \beta r s }|^2\approx \frac{(gg_A)^2m^2_Am^2}{(2m^2+2m^2_A-M^2)^2}.             \eea
\indent Therefore, the total thermal averaged cross-section for these annihilation processes reads

\begin{multline}
  %Simplificando   \langle \sigma v \rangle \sim \Big(  (\frac{\sqrt{m}}{\sqrt{M}})\frac{g^4T}{8\pi m^3} +\sum_i(\frac{\sqrt{m}}{\sqrt{m_i}})\frac{(gg_i)^2Tm_i}{2\pi(2m^2+2m^2_i-M^2)^2}+\sum_A(\frac{\sqrt{m}}{\sqrt{m_A}})\frac{(gg_A)^2m^2_A}{64\pi (2m^2+2m_A^2-M^2)^2}                      \Big)2\sqrt{\frac{2T}{m}} chegamos em 
       \langle \sigma v \rangle \sim \left(\frac{g^2T}{4 m^3 \sqrt{M}} +\sum_{\mathrm{J}}\frac{g_{\mathrm{J}}^2T\sqrt{m_{\mathrm{J}}}}{(2m^2+2m^2_{\mathrm{J}}-M^2)^2}+\sum_A\frac{g_A^2m^{3/2}_A}{32 (2m^2+2m_A^2-M^2)^2}                      \right)\frac{g^2}{\pi}\sqrt{{2T}} \end{multline}

It is possible to show that considering the range of DM masses and the constraint regarding the cosmological abundance, the dominant contribution comes from the Elko-vector boson scattering processes for the case of a DM with a weak scale mass. Then, considering this contribution for the prototype case \footnote{We leave for a forthcoming paper the full analysis of the whole allowed curve relating mass and coupling, see the last paragraph of this article.} $m \approx 100$ GeV, complying with a weak scale like mass range and (CDM), the cosmological abundance constraint defined by Eq.   \eqref{abund}  furnishes a first estimate on the order of magnitude of the coupling $g\approx 0.34$ being of the same order of the vector boson coupling.   \\
\indent As mentioned, the freeze-out temperature is obtained in the density stabilization threshold \footnote{$M_{pl}$ denotes the Planck mass.}, considering a non-relativistic dispersion at the decoupling, according to the cold dark matter (CDM) paradigm 
\bea 2\sigma \sqrt{\frac{2T}{m}} g^*\left(\frac{mT}{2\pi}\right)^{3/2}e^{-m/T}\approx \frac{\pi T^2 \sqrt{g_{eff}}}{3\sqrt{10}M_{pl}}.     \eea
 
\indent Considering this approximation, the cosmological abundance constraint,  and replacing the variables with their phenomenological values, yields \footnote{Ignoring the temperature dependence of $g_{eff}$, for simplicity \cite{Bauer:2017qwy}.}
\bea\label{ab1} e^{x_{\textit{eq}}}=4.44 \times 10^9x^{3/2}_{\textit{eq}}    \eea 
 Then, one obtains \footnote{ After using \textsc{Wolfram} to numerically find the root of Eq. (\ref{ab1}).}
\bea   x_{eq}\approx 27,\eea
ensuring the non-relativistic nature of the relic.\\
\indent Summarizing, we fixed our cross-section to be in agreement with the DM abundance. Later, we defined a reference value for the mass and coupling according to this constraint. Then, we calculated the decoupling temperature using the approximations mentioned in the excerpt between Eqs. (\ref{dmd}) and (\ref{dmd1}). The obtained value was indeed in compliance with our hypothesis of a non-relativistic/cold relic. Therefore, the issues on structure formation, inferred by model-independent cosmological observations assumed in the CDM model, are contemplated in our approach.\\

\section{Constraints from electron/DM scattering}\label{12}
\indent The proposed interaction is such that it allows a $2\to 2$ scattering between our DM candidate and electrons. Therefore, it is fundamental to verify whether our modeling of this process complies with the existing experimental bounds.  We consider recent data from Ref. \cite{ATLAS:2022gbw}, associated with electron-DM scattering from aromatic organic targets.\\
\indent The only channel contributing to this process is the $t$ one, with amplitude 
\bea i{\cal{M}}_{\alpha \beta \gamma' \sigma' }=-i\frac{gg_e}{m(t-M^2)} \gdualn{\xi}_\alpha(\p_1)\slashed{p}_2 \xi_\beta(\p_2) \Bar u_{\gamma'}^e(p_3) u_{\sigma'}^e(p_4).                   \eea
\noindent The non-polarized squared amplitude reads
\bea \frac{1}{8}\sum_{\alpha \beta \gamma' \sigma' }|{\cal{M}}_{\alpha \beta \gamma' \sigma' }|^2=\frac{(gg_e)^2}{8(t-M^2)^2}Tr[\slashed{p}_1\slashed{p}_2]Tr\Big[(\slashed{p}_3+m_e)(\slashed{p}_4+m_e)\Big].            \eea 
\indent Now, since the lab target of this process is placed on Earth approaching the center of galaxy \footnote{Here a significant amount of DM is supposed to exist. } with $v \approx 232\, {\rm Km/s}$, the relative velocity in natural units has order $v\sim 7.7 \times 10^{-4}$, leading to small contributions from momentum terms.
Going to the center-of-mass frame, one derives the electron and DM velocities $v_e\thickapprox \frac{m v}{m_e+m}$ and $v_\uplambda \thickapprox -\frac{m_p v}{m_p+m}$, respectively. \\
\indent The convenient variables for such a process are
\begin{subequations}
\bea    p_1^\mu&=&(E_\uplambda,0,0,p_\uplambda), \qquad\qquad\qquad\qquad p_2^\mu=(E_e,0,0,-p_\uplambda),\\
p_3^\mu&=&(E_\uplambda, p_\uplambda \sin \theta, 0, p_\uplambda \cos\theta), \qquad\qquad p_4^\mu=(E_e, -p_\uplambda \sin \theta, 0, -p_\uplambda \cos\theta),              \eea
\end{subequations}
with $E_e=\sqrt{p_\uplambda^2+m^2_e}$ and $E_\uplambda=\sqrt{p_\uplambda^2+m^2}$. Therefore, using $E_{CM}\approx m+m_{e}$ and $M\gg p_\uplambda$, one obtains the non-polarized squared amplitude disregarding momentum terms
\bea \frac{1}{8}\sum_{\alpha \beta \gamma'\sigma'}|\mathcal{M}_{\alpha \beta \gamma'\sigma'}|^2\approx 4(gg_{e})^2\frac{m^2 m^2_{e}}{M^4}. \eea
Therefore, the cross-section reads
\bea \sigma_{\uplambda e} \approx \frac{(gg_{e})^2m^2 m^2_{e}}{4\pi M^4(m+m_{e})^2}. \eea

\indent The cross-section $\Bar{\sigma}_{\uplambda e}$, suitable for our discussion, was  obtained in Ref. \cite{Blanco:2019lrf}
% (eq. 13) 
as
\bea  \frac{1}{16m^2m^2_e}\left(  \frac{1}{8}\sum_{\alpha \beta \gamma'\sigma'}|\mathcal{M}_{\alpha \beta \gamma'\sigma'}|^2 \right)=\frac{\pi \Bar{\sigma}_{\uplambda e} }{\mu^2}, \eea
with $\mu^2$ being the reduced mass of the system. We are considering the form factor equal to $1$, in compliance with our approximation. Hence, considering  $m\approx 100$ GeV and $g\approx 0.34$ yields\footnote{Considering the form factor is equal to the unit, due to the high magnitude of the Higgs mass intermediating the process.}
\bea \Bar{\sigma}_{\uplambda e} \approx 8\times 10^{-50}\,{\rm cm^2},  \eea
which is a value within the experimental bound derived in this reference.

The difficulty in directly detecting this process is due to the tiny value of Higgs-electron coupling and the fact that the highly massive Higgs particle enters as the mediator. This and the previous considerations reinforce the fact that the Higgs portal seems to be the most interesting one regarding the investigation of DM signatures. It provides a natural
explanation of the origin of some of the most important current phenomenological stringent limits. Namely, it supports the experimental bounds without the need to consider unnatural settings for the DM mass and the coupling $g$.\\
\indent Regarding the physical constraints and correlated bounds and estimations on the theory's parameters, it is important to clarify that we considered some reference approximate values, as a first consideration on orders of magnitude. Namely, for example, since the constraints are on the cross-sections and not directly on the mass and coupling constant, one could consider a whole curve of points labeled by these parameters and verify the allowed regions instead of just these reference values. 
However, taking into account the importance of the set of our achievements, they imply that cross-sections and other physical quantities have different functional dependencies on the parameters (as compared to the standard current approaches). It then opens the possibility of true new insights to describe dark matter pointing new directions for experimental searches, and correlated investigations on data analysis, including a wider consideration of graphical and computational tools having in mind a more precise definition of the allowed physical parametric regions. Moreover, the study of improved numerical solutions of the Boltzmann equations is also an interesting perspective. 

\section{Concluding Remarks}\label{13}

Throughout this paper, a variety of theoretical achievements were obtained as well as their main implications in terms of observables and phenomenology.
The self-conjugate and anti-self-conjugate Elko spinors were scrutinized, 
based on the Wigner degeneracy inducing the construction of a spinor dual which is compatible with the description of DM. Orthonormal relations and spin sums were consistently obtained for Elko spinors.
In this manner, the quantum mass dimension one fermionic field was appropriately constructed upon Elko spinors as expansion coefficients, using particle and anti-particle creation operators. The dual structure properly yielded a Hermitian free action for quantum Elko fields, with the additional possibility of a Hermitian interaction with the Higgs. In addition, we showed that the inherent properties of Elko fermionic fields evade any possibility of standard minimal couplings with electromagnetic fields and mediators of non-Abelian interactions. Other kinds of Hermitian couplings were also proposed. 
The Feynman rules for external Elko fields were addressed in the context of the interaction involving the Higgs, as well as discussions on 1-loop radiative corrections and renormalizability. Additionally, the M\o ller-like scattering at tree-level was investigated, with explicit computations regarding positive-defined squared amplitudes, in both polarized and non-polarized cases.

We also investigated several other channels regarding the annihilation of quantum Elko fermionic fields into Higgs particles, leptons, and vector bosons. These processes have a direct implication on the DM abundance and relic formation. Then, considering the cosmological constraint, a relation between the theory's free parameters was defined, taking into account a weak scale mass to ensure a (CDM) scenario. It also provided a suitable setup to discuss DM-electron scattering experiments. Interestingly, due to the tiny Higgs-electron coupling, the cross-section for the latter process is immediately within the experimental bounds. 

Summing up, although mainly focused on theoretical perennial improvements based on the last achievements on Wigner degeneracy, this work provides the link between these properties and phenomenological outputs placing mass dimension one fermions as  legitimate DM  candidates.

\subsection*{Acknowledgment}
The authors thank Professor Dharam Vir Ahluwalia (in memoriam) for always guiding and motivating our investigations in the field of Elko spinors. We also thank Marco Dias for reading the preliminary version of this work and for the discussions throughout the development of the research.
GBdG thanks the S\~ao Paulo Research Foundation FAPESP (2021/12126-5) for the financial support. 
RdR~is grateful to FAPESP (No. 2022/01734-7, and 	No. 2024/05676-7) and the National Council for Scientific and Technological Development -- CNPq (Grants No. 303742/2023-2 and No. 401567/2023-0), for partial financial support. RJBR thanks the generous hospitality offered by UNIFAAT and to Prof.  Renato Medina.

\appendix
\section{Phenomenology with arbitrary trial parameters}\label{a1}
Computations developed along the work were performed under very specific parameter fixation. We reserved this section to explicitly show the general counterpart of the amplitudes, with open factors associated with these parameters. In such case, the quantities given in Eqs. (\ref{ampli111}) -- \eqref{ampli3434}, yields
\begin{eqnarray}
 |{\cal{M}}_{1,1,1,1}|^2 = \left(
\begin{array}{c}
 \frac{g^4 M^4 p^4 \sin ^4(\theta ) (m+2 E )^4 (p^2-E^2)^2(t-u)^2 \left(\beta \alpha^*-\alpha \beta^*\right)^4 \left((m+E)^2-p^2\right)^2 \left(-M^2+t+u\right)^2}{16 m^{12} (m+E )^4 \left(M^2-t\right)^2 \left(M^2-u\right)^2} \\
\end{array}
\right) \nonumber \\
\end{eqnarray}

\begin{eqnarray}
|{\cal{M}}_{1,1,2,2}|^2 =    \left(
\begin{array}{c}
 \frac{g^4 M^4 p^4 \sin ^4(\theta ) (m+2 E )^4 (p+E )^4 (t-u)^2 \left(\beta \alpha^*-\alpha \beta^*\right)^4 (m+p+E )^4 \left(-M^2+t+u\right)^2}{16 m^{12}  (m+E )^4 \left(M^2-t\right)^2 \left(M^2-u\right)^2} \\
\end{array}
\right)
\end{eqnarray}

\begin{eqnarray}
&&|{\cal{M}}_{3,4,3,4}|^2 =   \nonumber 
\\
&&{\small\begin{array}{c}
 \frac{g^4 M^4 (p^2\!-\!E^2)^2(t\!-\!u)^2 \left(\beta \alpha^*\!-\!\alpha \beta^*\right)^4 \left((m\!+\!E)^2\!-\!p^2\right)^2  \left(t+u-\!M^2\right)^2 \left(2m^4\!-\! p^2 \cos (2 \theta ) \left(5 m^2\!+\!4 mE \!+\!4 p^2\right)+{m^2 p^2}\!+\!4 m E ^3\!+\!2p^4\!+\!2E ^4\right)^2}{32 m^{12}  (m+E )^4 \left(M^2-t\right)^2 \left(M^2-u\right)^2} 
\end{array}}\nonumber\\
\end{eqnarray}
%(a+ib)(c-id)-(c+id)(a-ib)=ac+bd+ibc-ida-ac-ad+ibc+bd=2bd

\indent Interestingly, for the case of real trial parameters, those squared amplitudes identically vanish. Therefore, it is zero for the case of the momentum space Majorana spinors \cite{Kirchbach:2003sa}.
%!! ref arXiv:hep-ph/0310297v2. Majorana def. eq. 6 and Elko em eq. 57. via fixing phases.

\end{document}